\documentclass[prl,twocolumn,showpacs,preprintnumbers,amsmath,amssymb]{revtex4}
\usepackage{graphicx,hyperref}
\usepackage{bm}

\newcommand\vp{\varphi}

\newcommand\Min{{\hbox{\begin{tiny}Min\end{tiny}}}}

\newcommand\ie{{\it i.e. }}

\newcommand\vpu{\vec\varphi_1}
\newcommand\vpd{\vec\varphi_2}
\newcommand\tr{\text{Tr}}

\DeclareRobustCommand\openone{\leavevmode\hbox{\small1\normalsize\kern-.33em1}}%

\def\nbC{{\mathchoice {\setbox0=\hbox{$\displaystyle\rm C$}%
\hbox{\hbox to0pt{\kern0.4\wd0\vrule height0.9\ht0\hss}\box0}}
{\setbox0=\hbox{$\textstyle\rm C$}\hbox{\hbox to0pt{\kern0.4\wd0\vrule
height0.9\ht0\hss}\box0}} {\setbox0=\hbox{$\scriptstyle\rm
C$}\hbox{\hbox to0pt{\kern0.4\wd0\vrule height0.9\ht0\hss}\box0}}
{\setbox0=\hbox{$\scriptscriptstyle\rm C$}\hbox{\hbox
to0pt{\kern0.4\wd0\vrule height0.9\ht0\hss}\box0}}}}
\def\nbQ{{\mathchoice {\setbox0=\hbox{$\displaystyle\rm
Q$}\hbox{\raise 0.15\ht0\hbox to0pt{\kern0.4\wd0\vrule
height0.8\ht0\hss}\box0}} {\setbox0=\hbox{$\textstyle\rm
Q$}\hbox{\raise 0.15\ht0\hbox to0pt{\kern0.4\wd0\vrule
height0.8\ht0\hss}\box0}} {\setbox0=\hbox{$\scriptstyle\rm
Q$}\hbox{\raise 0.15\ht0\hbox to0pt{\kern0.4\wd0\vrule
height0.7\ht0\hss}\box0}} {\setbox0=\hbox{$\scriptscriptstyle\rm
Q$}\hbox{\raise 0.15\ht0\hbox to0pt{\kern0.4\wd0\vrule
height0.7\ht0\hss}\box0}}}}
\def\nbT{{\mathchoice {\setbox0=\hbox{$\displaystyle\rm T$}\hbox{\hbox
to0pt{\kern0.3\wd0\vrule height0.9\ht0\hss}\box0}}
{\setbox0=\hbox{$\textstyle\rm T$}\hbox{\hbox to0pt{\kern0.3\wd0\vrule
height0.9\ht0\hss}\box0}} {\setbox0=\hbox{$\scriptstyle\rm
T$}\hbox{\hbox to0pt{\kern0.3\wd0\vrule height0.9\ht0\hss}\box0}}
{\setbox0=\hbox{$\scriptscriptstyle\rm T$}\hbox{\hbox
to0pt{\kern0.3\wd0\vrule height0.9\ht0\hss}\box0}}}}
\def\nbS{{\mathchoice {\setbox0=\hbox{$\displaystyle \rm
S$}\hbox{\raise0.5\ht0%  
\hbox to0pt{\kern0.35\wd0\vrule height0.45\ht0\hss}\hbox
to0pt{\kern0.55\wd0\vrule height0.5\ht0\hss}\box0}}
{\setbox0=\hbox{$\textstyle \rm S$}\hbox{\raise0.5\ht0% 
\hbox to0pt{\kern0.35\wd0\vrule height0.45\ht0\hss}\hbox
to0pt{\kern0.55\wd0\vrule height0.5\ht0\hss}\box0}}
{\setbox0=\hbox{$\scriptstyle \rm S$}\hbox{\raise0.5\ht0% 
\hboxto0pt{\kern0.35\wd0\vrule height0.45\ht0\hss}\raise0.05\ht0% 
\hbox to0pt{\kern0.5\wd0\vrule height0.45\ht0\hss}\box0}}
{\setbox0=\hbox{$\scriptscriptstyle\rm S$}\hbox{\raise0.5\ht0% 
\hboxto0pt{\kern0.4\wd0\vrule height0.45\ht0\hss}\raise0.05\ht0% 
\hbox to0pt{\kern0.55\wd0\vrule height0.45\ht0\hss}\box0}}}}
\def\nbZ{{\mathchoice {\hbox{$\sf\textstyle Z\kern-0.4em Z$}}
{\hbox{$\sf\textstyle Z\kern-0.4em Z$}} {\hbox{$\sf\scriptstyle
Z\kern-0.3em Z$}} {\hbox{$\sf\scriptscriptstyle Z\kern-0.2em Z$}}}}
\begin{document}

\title{Second order phase transitions induced by disorder in
frustrated magnets}

\author{Matthieu Tissier} \email{tissier@thphys.uni-heidelberg.de}
 \affiliation{Institut f\"ur Theoretische Physik, Universit\"at
 Heidelberg, Philosophenweg 16, 69120 Heidelberg, Germany}

%\date{\today}

\begin{abstract}
We study the critical properties of three dimensional frustrated
magnets, diluted with non-magnetic impurities. We show that these
systems exhibit a second order phase transition, corresponding to a
new universality class. In the pure case, the phase transition is
expected to be weakly of first order. We therefore argue that these
frustrated systems can be used to study experimentally the rounding
effect of disorder on discontinuous phase transitions. We give first
estimates of the critical exponents associated with this universality
class, by using the method of the effective average action.
\end{abstract}

\pacs{11.10.Hi, 75.40.Cx}

\maketitle

The influence of disorder on the properties of a system around a phase
transition has triggered a lot of studies in the last twenty years.
In the case of second order phase transitions, we have a fairly good
understanding of the situation thanks to the Harris criterion
\cite{harris74,chayes86}, which states that for a broad class of
systems with quenched disorder, the critical exponent $\nu$
(describing the singularity of the correlation length at the phase
transition) must satisfy the inequality: $\nu\geq 2/d$.  This has a
dramatic consequence for these pure materials for which the preceding
inequality is not satisfied, since a small amount of disorder changes
the critical exponents, such that the preceding inequality is
fulfilled. The pure system is therefore unstable with respect to the
inclusion of impurities, and one can expect a new universality class
associated with the disordered system.

Although much less studied, the influence of disorder on materials
undergoing first order phase transitions is also of great interest. In
such a situation, the general trend is that quenched disorder
smoothens the phase transition. In particular, Aizenman and Wehr
\cite{aizenman89} have proven that when disorder is coupled to a
system of small enough dimensionality \footnote{Here, small enough
means $d\leq4$ for systems with continuous order parameters (such as
$XY$ or Heisenberg spin systems), and $d\leq2$ for discontinuous order
parameters (such as Ising and Potts models)}, the density conjugated
to the disorder is a continuous function of the external parameters
(temperature, magnetic field, etc.). For a system with random bonds
(\ie a random distribution of the interaction between nearest
neighbors degrees of freedom), this statement implies that the energy
density, which is then the density conjugated to the disorder, is a
continuous function of the temperature. As a consequence, the latent
heat vanishes: the transition is indeed smoothened by the disorder.

This rounding effect of the disorder has been also studied in specific
models, mainly the random bond $q$-states Potts model. In two
dimensions, Monte-Carlo simulations, as well as theoretical work show
that the first order phase transition observed in the pure system for
$q\geq 5$ becomes of second order when quenched disorder is introduced
(see \cite{chatelain01} and references therein). Monte-Carlo
simulations on the three dimensional 3-states Potts model
\cite{ballesteros00} display the same feature. However a shortcoming
of the random bond Potts model for studying the rounding effect of the
disorder, is that it seems difficult to find experimental realizations
in the relevant cases. Indeed, for three-dimensional realizations of
the Potts model, the disorder usually induces a random field term, and
thus the relevant model is {\em not} the random bond Potts model
\cite{ballesteros00}. On the other hand, in $d=2$, it is required to
find an experimental realization of the Potts model with at least five
states, which seems quite difficult.

While studying the rounding effect of the disorder, it would be
valuable to find an experimental realization of this situation. With
this goal in mind, we consider in this letter the influence of
non-magnetic impurities in $XY$ and Heisenberg stacked triangular
antiferromagnets (STA), such as CsMnBr$_3$, CsNiCl$_3$, and rare earth
helimagnets (Ho, Tb, Dy). The spin degrees of freedom of these
magnetic materials are submitted to competing interactions, which make
the ground state more degenerate than in a non-frustrated system. For
the pure case, it was soon proposed that a new universality class
should be associated with the phase transition observed in these
materials \cite{bak76,yosefin85}, and from then on, there has been
numerous works aimed at characterizing it. We have now strong
evidences that this phase transition is actually weakly of first
order, and not continuous as was initially proposed. Let us indicate
the most striking evidences for this behavior in the $XY$ case. First,
most experimental and numerical studies indicate power-law behaviors
of the physical observables around the transition temperature,
characterized by a set of critical exponents. A detailed analysis
shows that these critical exponents are most probably not compatible
with those expected in a second order phase transition
\cite{tissier01}. It is more likely that the phase transition is
actually of first order, with such a large correlation length that the
discontinuity at the phase transition cannot be observed in
experiments. This scenario is confirmed by a recent Monte-Carlo
simulation \cite{itakura01} in which a first order phase transition
was directly observed, but only for very large lattice sizes. Finally,
a renormalization-group calculation in the framework of the effective
average action shows a lack of a stable fixed point \cite{tissier00}
for both $XY$ and Heisenberg cases, which in turn indicates that the
phase transition is of first order. Note however that six-loop
resummed series indicate a true second order phase transition
\cite{pelissetto01}.

At the light of the introductory discussion, we propose to study the
influence of quenched non-magnetic impurities in the critical behavior
of STA and helimagnets. In particular, we want to show that in
presence of impurities, the phase transition becomes continuous. We
also want to characterize the associated universality class by giving
a first estimate of the critical exponents. But before embarking on
this discussion, let us stress on two evidences which indicate the
importance of disorder in these systems. The first observation is
that, in the pure case, the effective exponent $\nu$ ($0.54\leq\nu\leq
0.57$ \cite{collins97}) is found to be much smaller than the lower
bound $2/3$ given by the Harris criterion ($\nu<2/d$ for disordered
systems), so that we can expect the disorder to be strongly relevant.
This conclusion is however an indication rather than a proof since
Harris criterion only applies to second order phase transitions, which
is not the case here. Second, the instability of the critical
properties of the system when impurities are added has been tested
experimentally on STA with diluted non-magnetic impurities
\cite{deutschmann92}, which shows, using scaling relation, that the
exponent $\nu$ increases strongly when impurities are introduced.

%\section{Hubbard-Stratonovitch transform}
We now derive the action relevant for the materials under study. Let
us consider the general situation of a diluted system of spins on a
lattice, submitted to a two-body interaction, in vanishing magnetic
field. The corresponding Hamiltonian reads:
\begin{equation}
\mathcal H=-\frac 12\sum_{i,j}J_{i,j}\;\epsilon_i\epsilon_j\vec
S_i\vec S_j .
\label{hamiltonien}
\end{equation}
Here, $J_{i,j}$ represents the (translation invariant) interaction
between spins at lattice sites $i$ and $j$, and the spins $\vec S_i$
are $n$-component vectors normed to unity. In physical realizations of
STA and helimagnets, the number of spin components is taken to be two
or three, depending on the anisotropies of the material. The dilution
appears through the random variables $\epsilon_i$, which equal 0 if a
non-magnetic impurity lies on site $i$, and 1 otherwise. The
$\epsilon_i$ are chosen to be independent random variables with no
dynamics, so as to describe a situation where the impurities are
quenched. The associated partition function reads:
\begin{equation}
\mathcal Z=\int\mathcal D \vec S \exp \left(-\mathcal H\right) .
\end{equation}
To deal with the unit norm constraint of $\vec S_i$, we perform the
standard Hubbard-Stratonovitch transform (we follow the discussion of
\cite{cardy84}) and introduce the auxiliary field $\vec \psi$, so that
the partition function reads:
\begin{equation}
\!\mathcal Z= \int \mathcal D \vec S \mathcal D \vec \psi
\exp\left(-\frac12 \sum_{i,j} J^{-1}_{i,j} \vec \psi_i \vec \psi_j+
\sum_i\epsilon_i \vec S_i\vec\psi_i\right)
\end{equation}
The variables $\vec S_i$ at different lattice sites are now decoupled,
and the integral on $\vec S$ can be performed. One gets, up to an
irrelevant multiplicative constant:
\begin{equation}
\mathcal Z=\int\mathcal D\vec\vp \exp\left(-\frac12 \sum_{i,j}
J^{-1}_{i,j} \vec \psi_i \vec \psi_j- \sum_i \epsilon_i
G_n(\vec\psi_i)\right)
\label{intermediaire}
\end{equation}
where \footnote{For $n=1$ (Ising model), $G_1(\vec x)=-\ln
(\cosh(|\vec x|))$}:
\begin{equation}
\label{def_fn}
G_n(\vec x)=-\ln\left(\frac{\int d^n\vec y \;\delta(\vec
y^2-1)\exp(\vec x.  \vec y)}{\int d^n\vec y \;\delta(\vec
y^2-1)}\right) ,
\end{equation}
For deriving Eq.(\ref{intermediaire}) we used the fact that the
$\epsilon_i$ take values in $\{0,1\}$.  In terms of the local
magnetization $\vec \vp_i=\sum_j J_{i,j}^{-1} \vec\psi_j$
\cite{cardy84}, the partition function reads:
\begin{equation}
\!\mathcal Z=\!\int\mathcal D\vec\vp \;\exp\left(\displaystyle
-\frac12 \sum_{i,j} J_{i,j} \vec \vp_i \vec \vp_j- \sum_i \epsilon_i
G_n\big({\textstyle \sum_j }J_{i,j}\vec\vp_j\big)\right)
\end{equation} 
The important point here is that $G_n(\vec x)$ is a scalar under
rotations (see Eq. (\ref{def_fn})) and therefore depends only on $\vec
x^2$. We conclude that the disorder $\epsilon_i$ couples to the
microscopic field $\vec\vp$ through terms with even powers of $\vec
\vp$ such as, for instance, the random mass term $\epsilon _i
\:\vec\vp_i^2$, but {\em not} through terms linear in the field, which
would correspond to the more complex situation of a random field.

We now specialize to the case of helimagnets and STA. In these
materials, the spin degrees of freedom are submitted to competing
interactions, which induce frustration. As a consequence, there are
two critical modes $\vec \vp_{\pm \vec Q}$ (\ie modes minimizing the
two-point interaction $J(\vec q)$ in Fourier space), with $\vec Q$
some characteristic momentum \cite{bak76,yosefin85}, instead of only
one critical mode $\vec \vp_{\vec q=0}$ in the ferromagnetic case. In
the long distance limit, the physics is governed by those modes whose
momenta are very close to $\pm \vec Q$, while the other modes only
influence non-universal quantities, and can be discarded. We are
therefore lent to introduce two types of fields:
\begin{equation}
\vp_\pm(\vec q)=\vp(\pm \vec Q+\vec q).
\end{equation}
It is convenient to perform the change of variables:
\begin{equation}
\begin{aligned}
\vp_1(q)&=\vec\vp_+(q)+i \vec \vp_-(q)\\ \vp_2(q)&=\vec\vp_+(q)-i \vec
\vp_-(q)
\end{aligned}
\end{equation}
If we now expand the interaction potential $G_n$ to fourth order in
the fields, the partition function can be recast into the form:
\begin{equation}
\mathcal Z=\int\mathcal D \vec\vp_1 \mathcal D \vec\vp_2\exp(-\mathcal
S)
\end{equation}
with:
\begin{equation}
\begin{split}
\mathcal S=\int d^dx\Bigg\{\frac{Z(x)}2\tr&\left(\partial
^t\phi.\partial \phi\right)+\frac{r(x)}2\rho+\\ &+\frac{g_1(x)}8
\rho^2+\frac{g_2(x)}4 \tau\Bigg\},
\end{split}
\label{action_desordre}
\end{equation}
We have merged here the two $n$-components vectors $\vpu$ and $\vpd$
into a $2\times n$ matrix $\phi=\left(\vpu,\vpd\right)$, and
introduced $\rho=\tr \left(^t\phi.\phi\right)$, $\tau=
\tr\left(^t\phi.\phi.^t\phi.\phi\right)$.  This action is very similar
to the one obtained in the pure case \cite{bak76,yosefin85}, except
for the $x$-dependence of the field normalization $Z$ and of the
coupling constants $r$, $g_1$ and $g_2$, appearing through the
presence of impurities (noted $\epsilon_i$ in
Eq.(\ref{hamiltonien})). In the following, we consider a direct
generalization of this situation where $\phi$ is a $m\times n$
matrix. For $m=2$, we retrieve the case of physical interest discussed
previously.

%\section{Average over disorder}
Once the action (\ref{action_desordre}) is determined, one would have
to compute the quantity of interest for a given realization of the
disorder (for a particular choice of the $\{\epsilon_i\}$). This is
however a formidable task. A simpler strategy consists in studying
self-averaging quantities averaged over the disorder. By doing so, one
recovers translational invariance. We use here the replica trick for
computing the average of the free energy over the disorder. We write:
\begin{equation}
\overline F=\overline{\ln \mathcal Z}=\lim_{o\rightarrow
0}\overline{\frac {\mathcal Z^o-1}o},
\end{equation}
where the line over a quantity represents the average with respect to
the disorder. The $o^{\text{\scriptsize th}}$ power of the partition
function can be written by introducing $o$ copies of the field
$\phi$. The average over disorder can now be computed (at least for a
Gaussian noise), and if we only keep the marginal and relevant terms
in the sense of renormalization group, we find the following
``$\phi^4$-like'' action:
\begin{equation}
\label{action_phi_4}
\begin{split}
\mathcal S_o(\phi_k)=\int &d^dx\Bigg\{\sum_{l=1}^o\Big[\frac Z2\tr
\left(\partial ^{\:t}\!\phi_l.\partial\phi_l\right)+r
\;\frac{\rho_l}2+\\&+\frac {u_1}8 \rho_l^2 +\frac{u_2}4\tau_l
\Big]+\frac{u_3}8\Big[\sum_{l=1}^o \rho_l\Big]^2\Bigg\},
\end{split}
\end{equation}
where the index $l$ labels the $o$ copies of the field. The critical
properties of the materials under study are obtained by using the
previous action, and taking the limit $o\rightarrow 0$ at the end of
the calculation.

%\section{Computation of the critical exponents}
Various methods can now be used for studying the critical properties
of the system (expansion in coupling constants,
$\varepsilon$-expansion, etc.). We propose here an approach based on
the concept of effective average action. This is a well documented
topic (see \cite{berges00} for a review, and
\cite{tissier00,tissier02} for applications in closely related
problems), so we only present the main ideas here. This approach is
based on the study of the effective average action $\Gamma_k$, which
interpolates smoothly between the microscopic action $\mathcal S$
(when $k\rightarrow\infty$) and the Gibbs free energy $\Gamma$ (when
$k\rightarrow 0$). $\Gamma_k$ therefore connects smoothly the
macroscopic and microscopic descriptions of the system. For
intermediates values of $k$, $\Gamma_k$ is obtained by integrating
over the high momentum modes only, while long distance modes are
effectively left unintegrated. A major interest of $\Gamma_k$ is that
its evolution with the scale $k$ is governed by an exact, concise
equation~\cite{tetradis94}:
\begin{equation}
\label{flot_G}
\frac {\partial \Gamma_k}{\partial t}=\frac12\int_q\frac {\partial_t
R_k(q)}{\Gamma^{(2)}(q)+R_k(q)},
\end{equation}
with $k\propto \exp(t)$. The function $R_k(q)$ describes how the low
and high momentum modes are separated, and $\Gamma_k^{(2)}$ is the
second functional derivative of the effective average action.
Although this equation cannot be solved exactly, it can be used in
practice to determine nonperturbative flow equations. The idea
consists in considering a truncation for $\Gamma_k$, typically
characterized by a finite number of coupling constants. By introducing
this truncation in equation (\ref{flot_G}), one deduces flow equations
for the different coupling constants. These are non-polynomial in
coupling constants, as can be seen from the nonlinear structure of the
flow equation (\ref{flot_G}).

We now describe the type of truncations considered here:
\begin{equation}
\begin{split}
\label{troncation}
\Gamma_k=&\int_x \frac Z2\sum_l\tr\left(\partial
^{\:t}\!\phi_l.\partial\phi_l\right)+\frac{u_1}8 \sum_l\left(\rho_l-
\kappa \right)^2\\& + \frac{u_2}4\sum_l\left(\tau_l-2\rho_l
\kappa\right)+\frac{u_3}8\left(\sum_l (\rho_l -\kappa)\right)^2
\end{split}
\end{equation}
This truncation is to be seen as the first terms of an expansion of
the effective average action around a configuration $\phi_\Min$,
defined by $(\phi_{\Min\;i,j})_l=\sqrt\kappa \delta_{i,j}$ (here, $i$
and $j$, running respectively from 1 to $n$ and from 1 to $m$, stand
for the matrix indices of the field, while $l$ stands for the replica
index, running from 1 to $o$). We have also considered truncations
with $\phi^6$ terms added. The action depends then on six more
coupling constants.  The flow equations have been derived using the
method presented above. However those are to lengthy to be displayed
here. We only quote the flow equation in the limit of small coupling
constants $\{u_1,u_2,u_3\}$, which identify with the one loop
expansion in coupling constant \footnote{We thank P. Lecheminant for
comparison with his unpublished results.}:
\begin{align}
\dot{u_1} &=
\begin{aligned}[t]
-\varepsilon u_1 + 2v_4\Big( ( 8& + mn ) u_1^2 + 12u_2^2 +\\+& 4u_1 (
  ( 1 + m + n ) u_2 + 3u_3 ) \Big)
\end{aligned} 
\\ \dot{u_2}&=-\varepsilon u_2 + 4 v_4u_2\Big( 6(u_1 +u_3) + ( 4 + m +
n ) u_2 \Big) \\ \dot{u_3}&=-
\begin{aligned}[t]
\varepsilon u_3 + & 2 v_4u_3\Big( 2( 2 + mn ) u_1 +\\&+ 4( 1 + m + n )
  u_2 + ( 8 + mno ) u_3 \Big)
\end{aligned}
\end{align}
where $\varepsilon=4-d$, $v_4=1/32\pi^2$, and the dot indicates a
derivative with respect to $t$. For generic values of $m$, $n$ and
$o$, the flow equations admit eight fixed points. Out of them, six are
already well known. There are four fixed points with $u_2=0$,
corresponding to the situation of a diluted system of $m n$ components
spins \cite{lubensky75}, and two extra fixed points with $u_3=0$,
corresponding to the pure frustrated magnets \cite{bak76}. In
addition, there are two new fixed points. We finally indicate that for
certain values of $(m,n,o)$, there is a degeneracy of the one loop
$\beta$ functions (for example, for $m=2$, $o=0$, $n\simeq 25.3$). For
such values of $(m,n,o)$, we expect a $\sqrt{\varepsilon}$-epsilon
fixed point to appear at two-loop order, as in the random bond Ising
model \cite{lubensky75}. We postpone the discussion of this point, as
well as a more detailed description of the fixed points to a
forthcoming publication.

%\section{Results}

\begin{figure}[htbp] 
\begin{center}
\includegraphics[width=7cm]{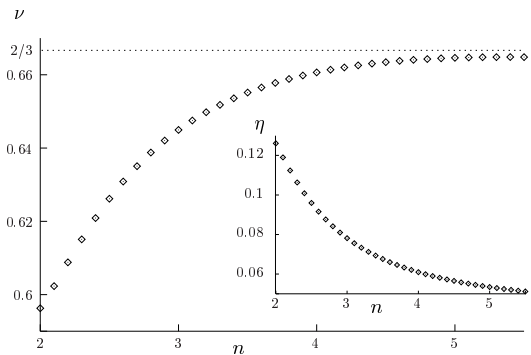}
\end{center}
\caption{Critical exponents of the diluted STA, as a function of the
number of spin components. The dotted line corresponds to the lower
bound given by Harris criterion.}
\label{exposants}
\end{figure}

In three dimensions, we have looked for roots of the $\beta$
functions. In the case of physical interest $m=2$, the disorder is
found to be irrelevant for $n\gtrsim 5.5$. On the other hand, for
$n\lesssim 5.5$, the disordered fixed point is stable and governs the
phase transition. We give on Fig.\ref{exposants} the values of the
associated critical exponents.
We first remark that our values of $\nu$ are all smaller than the
lower bound $2/3$ predicted by the Harris criterion. However, for
$3\leq n\leq 5.5$, $\nu$ is off by only few percents, which is
typically the error expected with the truncation considered here. The
much larger discrepancy for $n=2$ can be attributed to the unusually
large value of the anomalous dimension $\eta$. In this case, richer
truncations are to be considered in order to get more accurate results
(work in progress). Although the accuracy of our results is
questionable, in particular for $n<3$, they indicate that $\nu$ is
varying very slowly with $n$ (in the pure vectorial model, the
variation of $\nu$ on the same range of $n$ is almost ten times
bigger), and that $\nu$ is close to the lower bound $2/3$ imposed by
Harris criterion. These features are similar to what is observed in
$d=2$ $q$-state Potts model with random bond using Monte-Carlo
simulations, where $\nu$ is found to be very close to the lower bound
$1=2/d$ obtained from the Harris criterion, for all $q\geq 2$.

To conclude, we have highlighted some interesting properties
associated with the inclusion of non-magnetic impurities in
three-dimensional frustrated magnets. In particular, a new
universality class was shown to appear, with associated critical
exponents differing strongly from the effective exponents measured in
the pure case. We also gave some indications for a very appealing
behavior of the critical exponent $\nu$, and a large value of
$\eta$. A verification of these effects by independant approaches, in
particular perturbative expansions and Monte-Carlo simulations, would
be very valuable.  At the light of the results presented here, it
would be particularly interesting to reconsider the experimental
studies on diluted frustrated magnets.
\begin{acknowledgments}
We thank B. Delamotte, D. Mouhanna and M. Picco for useful
discussions. This work was supported by Marie Curie fellowship
HPMF-CT-2001-01343.
\end{acknowledgments}

\end{document}